\documentclass[review,hidelinks,onefignum,onetabnum]{siamart220329}

\usepackage{lipsum}
\usepackage{amsfonts}
\usepackage{graphicx}
\usepackage{epstopdf}
\usepackage{algorithmic}
\usepackage{amsopn}
\usepackage{amsmath}
\usepackage{amssymb}
\usepackage{mathrsfs}
\usepackage{bm}

\nolinenumbers
\title{Resilience patterns in higher-order meta-population networks
	\thanks{Submitted to the editors DATE.
		\funding{This work was partially supported by the Social Science Foundation of Chongqing, No. 2021PY53, Natural Science Foundation of Chongqing, No. cstc2021jcyj-msxmX0132, Natural Science Foundation of Yuzhong District, Chongqing, No. 20210117, and National Natural Science Foundation of China under Grants No. 61903266 and No. 62272074.}}\\(\emph{Corresponding author: Tao Lin}\\(\emph{Yanyi Nie and Wei Wang contribute equally.})}

\author{Yanyi Nie\thanks{College of Computer Science, Sichuan University, Chengdu 610065 (\email{nienienie@stu.scu.edu.cn}, \email{lintao@scu.edu.cn}).}
	\and Yanbing Liu\thanks{Chongqing Medical University, Chongqing, 400016, China (\email{liuyb@cqupt.edu.cn}).}
	\and Qixuan Cao\footnotemark[3]
		\and Tao Lin\footnotemark[2]
	\and Wei Wang \thanks{School of Public Health, Chongqing Medical University, Chongqing, 400016, China (\email{wwzqbx@hotmail.com}, \email{xuqxccool@163.com}).}
}
\ifpdf
\hypersetup{
  pdftitle={Resilience patterns in higher-order meta-population networks},
  pdfauthor={Yanyi Nie, Yanbing Liu, Qixuan Cao, Tao Lin,  Wei Wang}
}
\fi

\begin{document}

\maketitle

\begin{abstract}
Meta-population networks are effective tools for capturing population movement across distinct regions, but the assumption of well-mixed regions fails to capture the reality of population higher-order interactions. As a multidimensional system capturing mobility characteristics, meta-population networks are inherently complex and difficult to interpret when subjected to resilience analysis based on $N$-dimensional equations. We propose a higher-order meta-population model that captures large-scale global cross-regional mobility and small-scale higher-order interactions within regions. Remarkably, we extend the dimension-reduction approach, simplifying the $N$-dimensional higher-order meta-population system into a one-dimensional equation by decomposing different network behaviours into a single universal resilience function, thereby allowing for convenient and accurate prediction of the system resilience. The network structure and human mobility parameters can clearly and simply express the epidemic threshold. Numerical experimental results on both real networks and star networks confirm the accuracy of the proposed dimension-reduction framework in predicting the evolution of epidemic dynamics on higher-order meta-population networks. Additionally, higher-order interactions among populations are shown to lead to explosive growth in the epidemic infection size potentially. Population mobility causes changes in the spatial distribution of infectious diseases across different regions.
\end{abstract}



\section{Introduction}
Incorporating population mobility into models is crucial for accurately depicting the spread of global epidemics \cite{Nilforoshan2023Humana,Gonzalez2008Understandinga,Chang2021Mobilitya}. Structured meta-population models \cite{Colizza2007Invasionc,Soriano-Panos2022Modelingb} serve as an important tool for capturing spatial movements on a large scale. Generally, a meta-population is a collection of spatially separated subpopulations that interact through population mobility. Meta-populations provide a highly useful framework for integrating reaction-diffusion processes, which has led to their widespread application in the fields of epidemiology \cite{Colizza2007Reactionc,Allard2023Role}, information and social contagion \cite{Wang2017Interplaya,Nie2022Informationb} and cellular diffusion \cite{Gonzalez-Garcia2002Metapopulationa}. The ``diffusion stage" describes the cross-regional spatial movement on a macro-scale. Each individual can change their current location and move to another subpopulation, thereby facilitating the spread of epidemics at the systemic level. For instance, the meta-population is used to capture the bidirectional mobility on individual mobility networks \cite{Belik2011Naturald}, population density mobility \cite{Hazarie2021Interplayb}, spatial distribution of the population, commuting flow patterns, and the interactions between these factors and epidemic spread \cite{Gomez-Gardenes2018Criticala,Angstmann2021General}.
The ``reaction stage" focuses on the infection between susceptible hosts and infected individuals within a subpopulation on the micro-scale. Generally, metapopulation models assume that the population within each subpopulation is sufficiently large and well-mixed \cite{Soriano-Panos2018Spreadingc,Arenas2020Modelinge,Nie2023Pathogenc,Choi2022Identifying}, allowing the pathogen spreads to any healthy individual within the same subpopulation with equal probability. Based on the well-mixed assumption, the meta-population is used to study the geographic locations that trigger epidemics \cite{Soriano-Panos2018Spreadingc}, regional incidence rates \cite{Arenas2020Modelinge}, and pathogen diversity \cite{Nie2023Pathogenc}.


However, the metapopulation structure of some small-scale groups significantly differs from that of large-scale urban environments \cite{Kraft2023Metapopulation,Paternoster2020Epidemic}. Examples include streets or neighbourhoods based on administrative divisions and villages linked by kinship, visitation and trade. Defining the contact patterns within these small-scale groups with the assumption of well-mixed interactions may lead to serious biases in epidemic predictions. Indeed, existing research has highlighted the impact of contact network structure on epidemiological outcomes \cite{Onaga2017ConcurrencyInduceda,Pastor-Satorras2001Epidemici,Granell2013Dynamicalb}, leading to the development of numerous mathematical models capable of representing underlying simple pairwise network structures \cite{Pastor-Satorras2002Epidemica,Bansal2007When} and spatial organization \cite{Barthelemy2011Spatiala}. Furthermore, as real systems exhibit a large number of higher-order interactions, such as super-spreading events in social systems \cite{Mancastroppa2022Sidewardc}, co-authorship relationships in scientific networks \cite{Patania2017shapei}, and protein interactions in biological networks \cite{Estrada2018Centralitiesc}, higher-order networks, namely simplicial complexes \cite{Schaub2020Randomb,Battiston2020Networksk,Iacopini2019Simplicialh} and hypergraphs \cite{Lim2020Hodge,Zhang2023Higherorder,Bick2023What}, have been introduced to extend pairwise interactions of contact patterns to higher-order interaction paradigms and extensively applied in epidemiological research \cite{Wang2024Epidemicc}. For the classical susceptible-infected-susceptible (SIS) epidemic, higher-order interactions, based on simplicial complexes and hypergraphs, have been proven to fundamentally alter the outbreak patterns of contagion dynamics \cite{Kim2024HigherOrdera}, leading to the emergence of bistability, explosive transitions, and hysteresis loop phenomena \cite{Ma2024Impact,Skardal2019Abrupte,Li2021Contagionc}. For susceptible-infected-recovered (SIR) epidemics, by extending the simplicial contagion model to the simplicial susceptible-infectious-recovered (s-SIR) model, not only can the discontinuous transitions and bistability of complex systems be captured \cite{Li2022Twod,Chen2024SIQRSa}, but also the periodic phenomena of epidemic outbreaks can be detected \cite{Wang2021Simplicialb,Palafox-Castillo2022Stochasticc}.

A metapopulation with nested reaction-diffusion processes is a dynamical system \cite{Brin2002Introduction,Close2001Modeling}. Resilience analysis \cite{Artime2024Robustnessa,Perrings1998Resilience,May1977Thresholds} quantitatively characterizes the ability of a dynamical system to recover from failures or malfunctions in network components (nodes), providing a scientific basis and guidance for adjusting its activities to maintain basic functions and avoid collapse. Traditional resilience analysis tools are usually conveniently applied to low-dimensional systems. For example, percolation theory-based methods analyze low-dimensional internet networks by determining the critical proportion of nodes that need to be removed before the network disintegrates, revealing the extremely high resilience exhibited by the internet \cite{Cohen2000Resiliencea,Laishram2018Measuring}. Besides, dynamical systems can be mathematically approximated to a one-dimensional nonlinear dynamical equation, and the system's resilience function is derived by calculating the fixed points and the second derivative of the equation, which is then used to analyze the system resilience \cite{Iacopini2019Simplicialh,Jana2016Optimal,Zhang2020Epidemic}. There are two entry points for complex multi-dimensional systems for resilience analysis methods. The first is to analyze the resilience of multi-dimensional systems directly. For instance, percolation theory is a universal method for analyzing the resilience of networks with community structures \cite{Dong2018Resilience,Shang2020Generalized,Ma2022Enhancing} and interdependent networks \cite{Dong2021Optimal,Shekhtman2015Resilience,Stippinger2014Enhancing}, that evaluates and predicts the overall robustness and recovery capacity of the system by studying random failures and connectivity changes of nodes or edges within the network. The second approach utilizes the dimension-reduction approach proposed by Gao et al. \cite{Gao2016Universal}, which compresses multi-dimensional systems into one-dimensional systems by decomposing different network behaviours into a single general resilience function, and then traditional resilience tools are used for analysis \cite{Zhang2020Resilience,Su2018Optimal,Ghosh2023Dimensiona,Tu2017Collapse}. Metapopulation networks, while capturing the inherent complexity of multi-dimensional systems, also capture the characteristics of population mobility between multiple subpopulations, making resilience analysis quite challenging. Existing studies \cite{Soriano-Panos2022Modelingb,Hazarie2021Interplayb,Gomez-Gardenes2018Criticala} typically focus on the multi-dimensional attributes of metapopulations. Using $N$-dimensional equations to capture their internal dynamic processes and derive critical points concerning matrix eigenvalues makes the resilience characterization process of metapopulations very complex and difficult to interpret.

Overall, the above analysis indicates that the assumption of uniform mixing within metapopulation networks in existing studies fails to capture actual population interactions. Moreover, resilience analysis methods for metapopulations are exceedingly complex and difficult to interpret. Therefore, this work proposes a metapopulation model with higher-order interactions that can simultaneously capture large-scale population mobility between regions from a global perspective and higher-order interactions within regions from a local perspective. Even more impressively, by extending the dimension-reduction approach, the $N$-dimensional microscopic Markov equation set describing the higher-order metapopulation system is collapsed into a one-dimensional equation, providing a simple and accurate mathematical framework for predicting system resilience. Based on macroscopic resilience parameters, network structure and population mobility features accurately characterise the epidemic outbreak threshold. We conducted extensive numerical experiments using Monte Carlo simulations, and the analysis results on real networks and star networks revealed the accuracy of the proposed dimensionality reduction theoretical framework in predicting the evolution of epidemic dynamics. Furthermore, higher-order interactions among populations can potentially lead to explosive growth in epidemic infections. Population mobility causes changes in the spatial distribution of infectious diseases in different regions. There is an important interplay between higher-order interactions and population mobility, which influence the dynamics of epidemic spread.

The the paper is organized as follows. Section \ref{model} provides a detailed introduction to the metapopulation model with higher-order interactions. Section \ref{theoretical}, we first describe the diffusion-reaction process with higher-order interactions using the MMCA. We then extend the dimension-reduction approach to simplify the $N$-dimensional equation set into a one-dimensional equation and analyze system resilience. Section \ref{numerical} presents several numerical experiments. Section \ref{conclusion} concludes the study and outlines directions for future research.

%

\section{Model descriptions} \label{model}
In this section, we present the reaction-diffusion process in a higher-order meta-population. Firstly, we introduce the higher-order meta-population network in Sec.\ref{high-meta}. Subsequently, the reaction-diffusion process with simplicial contagion is detailed in Sec.\ref{r-d}.
\begin{figure*}[h]
	\centering
	\includegraphics[scale=0.47]{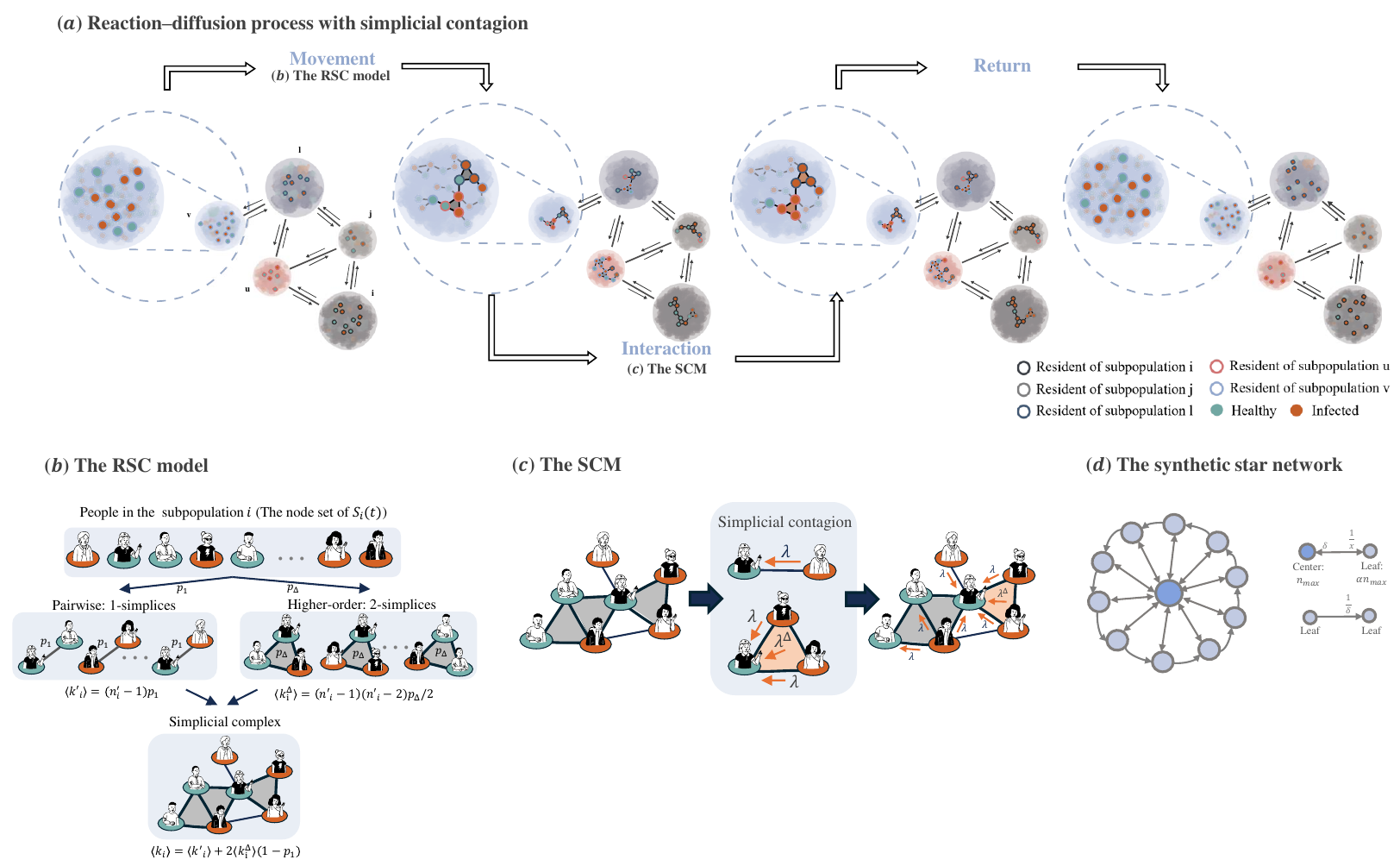}
	\caption{(Color online) The reaction-diffusion process with simplicial contagion in the higher-order meta-population model. \textbf{(a)} Four snapshots of the Movement-Interaction-Return process in the meta-population from a microscopic perspective. The first snapshot depicts the initial state of the meta-population with five subpopulations, $i$, $j$, $l$, $u$, and $v$. The five large circles are different subpopulations (nodes). The small circles within each subpupolation represent individuals with different residences and states. Referring to the legend in the bottom-right corner, the outer colour of the small circles corresponds to their residence, while the inner colour represents their state. An enlarged view of subpopulation $v$ is provided inside the blue dashed circle. The second snapshot shows the distribution of individuals in the meta-population after moving.  The third snapshot displays the distribution of individuals after infection and recovery occur. The fourth snapshot illustrates the outcome of individuals returning to their places of residence. \textbf{(b)}  Schematic illustration of the RSC model. \textbf{(c)} Schematic diagram of the SCM. \textbf{(d)} The synthetic star-shaped  network. The central node has a degree of $k$, and the population ratio between the central node and leaves is $\alpha$, with a mobility factor from leaves denoted as $\delta$. An individual's probability of moving from the central node to any leaf is $\frac{1}{x}$.}
	\label{fig0}
\end{figure*}

\subsection{Higher-order meta-population network}\label{high-meta}

A typical meta-population model consists of $N$ subpopulations (nodes) and a population of $M$ individuals. Each subpopulation $i\in[1,N]$ contains $n_i$ individuals who reside in subpopulation $i$. Thus, each individual associated with its corresponding residence $i$ implies $M=\sum_{i=1}^N n_i$.

The mobility of individuals is the most crucial feature of meta-population frameworks. Each individual in the system can change its current location, moving to another node, thereby facilitating the spread of epidemics at the system level. The mobility paths of individuals, connecting various nodes, are typically represented by a weighted directed graph encoded in an adjacency matrix $\rm{W}$. The entry $W_{ij}$ represents the weight of interaction, indicating the probability of moving from node $i$ to node $j$.

In a higher-order meta-population model, similar to a typical meta-population model, residents within each subpopulation (node) are initially well-mixed. Specifically, after mobility occurs, interactions among individuals within a subpopulation $i$ are no longer uniformly equivalent but are captured by a simplicial complex $\mathcal{S}_i$. A simplicial complex $\mathcal{S}$ is a common model that simultaneously represents pairwise and higher-order interactions, composed of a set of simplices. In essence, a $d$-simplex $\sigma$ is formed by $d+1$ distinct vertices, which can be represented as $\sigma = [v_0, v_1, \cdots, v_d]$. Following normal network nomenclature, simplices of different dimensions are classified as follows: a single vertex is termed a 0-simplex, an edge connecting two vertices is termed a 1-simplex, a ``full" triangle connected by three vertices is termed a 2-simplex, and so on. In particular, the characterization of the collective behaviour among three elements can be achieved in two different ways, i.e. a single 2-simplex ($[v0, v1, v2]$) or a set of three 1-simplices ($[v0, v1], [v0, v2], [v1, v2]$). Clearly, for any given simplicial complex $\mathcal{S}$, if simplex $\sigma$ belongs to $\mathcal{S}$, then all subsimplices $v\subset \sigma$ formed by the vertices of $\sigma$ are also included in $\mathcal{S}$.

\subsection{Reaction–diffusion process with simplicial contagion}\label{r-d}
The dynamics occurring within the higher-order metapopulation model are described by the reaction-diffusion process with simplicial contagion, which involves three distinct stages at time step $t$: movement, interaction, and return. 
Specifically, the description of the three stages involved in the dynamics is as follows:

(1) Movement. Firstly, as illustrated in Fig.~\ref{fig0}(a), each individual $v_i$, where $i\in[1,N]$, needs to decide whether to move with probability $p_d$ or to stay in its residential node $i$ with probability $(1-p_d)$.

Case 1: If $v_i$ moves with probability $p_d$, then according to the adjacency matrix $\mathbf{W}$, it moves to any neighbouring subpopulation of node $i$. Specifically, the probability of node $v_i$ moving to subpopulation $j$ is represented by $R_{ij}$, expressed as
\begin{equation}
	R_{ij}=\frac{W_{ij}}{\sum_{j=1} ^N W_{ij}},
\end{equation} 
which is proportional to the corresponding entry $W_{ij}$ of the adjacency matrix, i.e., the interaction weight.

Case 2: If $v_i$ stays with probability $1-p_d$, $v_i$ remains in the current subpopulation $i$.

Once the movement is completed, at time $t$, the higher-order contact network among individuals within the same subpopulation $i$ is described by the simplicial complex $\mathscr{S}_i(t)$. The construction of the simplicial complex is described by the random simplicial complex (RSC) model \cite{Gomez-Gardenes2018Criticala} as illustrated in Fig.~\ref{fig0}(b), with the following specific steps.

(i) At time $t$, all individuals located in subpopulation $i$ (residency not necessarily $i$) constitute the node set of the simplicial complex $\mathscr{S}_i(t)$, with a size of $n_i'(t)$.

(ii) Given the fixed average degrees $\langle k_i \rangle$ and $\langle k_i^{\Delta} \rangle$ of the simplicial complex $\mathscr{S}_i(t)$ ($t=0,\cdots,\infty$), and based on the formulas $p_1=\frac{\langle k_i \rangle-2\langle k_i^{\Delta} \rangle}{[n_i'(t)-1]-2\langle k_i^{\Delta} \rangle}$ and $p2=\frac{2\langle k_i^{\Delta} \rangle}{[n_i'(t)-1][n_i'(t)-2]}$ \cite{Gomez-Gardenes2018Criticala}, calculate the probabilities $p_1$ and $p_{\Delta}$ for constructing 1-simplices and 2-simplices in the simplicial complex $\mathscr{S}_i(t)$.

(iii)  Referring to the RSC model, firstly connect any pair of individuals $(v_i,v_j)$ with probability $p_1$ to construct a 1-simplex. At this stage, each individual's average degree of 1-simplices is $k'_i=[n_i'(t)-1]p_1$.
Then, with probability $p_{\Delta}$, connect any triplet $(v_i,v_j,v_l)$ to construct a 2-simplex. After this step, the average degree of 2-simplices for each individual is $k_i^{\Delta}=[n_i'(t)-1][n_i'(t)-2]p_{\Delta}/2$. Additionally, the construction of 2-simplices contributes to increasing each individual's average degree of 1-simplices by $2k_i^{\Delta}(1-p_1)$. Hence the final average degree of the simplicial complex $\mathscr{S}_i(t)$ is $\langle k_i \rangle=\langle k'_i \rangle+2k_i^{\Delta}(1-p_1)$.  In our model, we limit the RSC model to dimension $D=2$.

Upon the completion of the RSC model within each subpopulation at time $t$, the resultant higher-order contact networks can be articulated as a set of simplicial complexes, formally expressed as $\mathscr{S}(t)=\{\mathscr{S}_1(t),\mathscr{S}_2(t),...,\mathscr{S}_N(t)\}$.

(2) Interaction. Based on the constructed higher-order networks following mobility, denoted as $S(t)$, infection and recovery occur within each subpopulation as shown in Fig.~\ref{fig0}(a). For individuals in any subpopulation $i$, their infection and recovery processes adhere to the simplicial susceptible–infected–susceptible (s-SIS) model, detailed as follows.

As illustrated in Fig.~\ref{fig0}(c), the infection process is delineated according to the simplicial contagion model (SCM). Within any simplicial complex $\mathscr{S}_i(t)$, the infection among individuals propagates through 1-simplices or 2-simplices.

In a 1-simplex, when there exists an $\mathcal{S}$-state (healthy) individual $v_i$ and an $\mathcal{I}$-state (infected) individual $v_j$, the $\mathcal{S}$-state individual becomes infected and transitions to $\mathcal{I}$-state with probability $\lambda$ through the 1-simplex. Such processes can be represented as
$$
\mathcal{S}+\mathcal{I} \xrightarrow[v_j\rightarrow v_i]{\lambda}\mathcal{I}+\mathcal{I}.
$$

In a 2-simplex, infection occurs in two scenarios when the $\mathcal{S}$-state individual $v_i$ is connected to two other individuals $v_j$ and $v_l$.

Case 1: Assuming only one of the individuals $v_j$ or $v_l$ is in $\mathcal{I}$-state, infection propagates through the 1-simplex formed by the link $v_j \rightarrow v_i$ or $v_l \rightarrow v_i$ with probability $\lambda$. The $\mathcal{S}$-state individual becomes infected and transitions to $\mathcal{I}$-state with probability $\lambda$, which can be expressed as
$$
\mathcal{S}+I \xrightarrow[v_j/v_l\rightarrow v_i]{\lambda}\mathcal{I}+\mathcal{I}.
$$

Case 2: Assuming both individuals $v_j$ and $v_l$ are in $\mathcal{I}$-state, infection propagates not only through the 1-simplices formed by the links $v_j \rightarrow v_i$ and $v_l \rightarrow v_i$ but also through the 2-simplex with probability $\lambda^{\Delta}$. The $\mathcal{S}$-state individual becomes infected and transitions to $\mathcal{I}$-state with probability $1-(1-\lambda)^2(1-\lambda^{\Delta})$. The process can be represented as
$$
\mathcal{S}+\mathcal{I}+\mathcal{I} \xrightarrow[v_j+v_l\rightarrow v_i]{1-(1-\lambda)^2(1-\lambda^{\Delta})}\mathcal{I}+\mathcal{I}+\mathcal{I}.
$$

Besides, based on the s-SIS model, any $\mathcal{I}$-state individual spontaneously recovers to $\mathcal{S}$-state with probability $\mu$, read as
$$
\mathcal{I} \xrightarrow{\mu}\mathcal{S}.
$$

(3) Return. Finally, as shown in Fig.~\ref{fig0}(a), all individuals return to their corresponding residential subpopulations due to the commuting nature of displacements. Individuals within each subpopulation revert to well-mixed and await the reaction-diffusion process in the next step.

\section{Theoretical analysis}\label{theoretical}
In this section, we first extend the microscopic Markov chain approach (MMCA) in Sec.\ref{muliti}, relying on a closed set of equations in $N$ dimensions to comprehensively describe the reaction-diffusion process with simplicial contagion. Then, in Sec.\ref{resi}, we analyze the resilience of higher-order meta-population systems by quantifying the average dynamics of neighbour nodes, reducing the dimensionality of the evolution equations, and computing analytical solutions for epidemic thresholds.

\subsection{Multidimensional evolution equations} \label{muliti}
Given $N$ subpopulations (nodes), $\rho_{i}^{I}(t)$ $(i=1,\cdots,N)$ defines the proportion of individuals in  $\mathcal{I}$-state residing in node $i$ at time $t$. The time evolution of $\rho_{i}^{I}(t)$ is
\begin{equation}
	\label{eq.2}
	\rho_{i}^{I}(t+1)=(1-\mu)\rho_{i}^{I}(t)+[1-\rho_{i}^{I}(t)]\Pi_{i}(t).
\end{equation} 
The first term on the right-hand side of Eq.~(\ref{eq.2}) represents the proportion of $\mathcal{I}$-state individuals residing in node $i$ that do not recover at time $t+1$. The second term, $[1-\rho_{i}^{I}(t)]\Pi_{i}(t)$, explains the proportion of $\mathcal{S}$-state individuals residing in node $i$ that are infected and transform into $\mathcal{I}$-state at time $t+1$. The $\Pi_{i}(t)$ is interpreted as the probability that individuals residing in node $i$ are infected at time $t$, expressed as
\begin{equation}
	\label{eq.3}
	\Pi_{i}(t)=(1-p_d)Q_{i}(t)+p_d\sum_{j=1}^N\frac{W_{ij}}{\sum_{l=1}^NW_{il}}Q_{j}(t),
\end{equation} 
where $p_d$ denotes the probability of mobility, while $W_{ij}$ is the weight of the link between nodes $i$ and $j$. The first term on the right-hand side indicates the probability of individuals being infected while remaining at node $i$. The second one considers the probability of individuals being infected when moving to any neighbour of node $i$. The $Q_{i}(t)$ represents the probability of individuals in (but not necessarily residing in) node $i$ are infected by other $\mathcal{I}$-state individuals within the same node $i$ at time $t$. The probability $Q_{i}(t)$ is written as
\begin{equation}
	\label{eq.4}
	Q_{i}(t)=1-(1-\lambda  \frac{I_i^{\rm eff}(t)}{n_i^{\rm eff}})^{\langle k_i \rangle}\big(1-\lambda^{\Delta} [\frac{I_i^{\rm eff}(t)}{n_i^{\rm eff}}]^2\big)^{\langle k_i^{\Delta} \rangle},
\end{equation} 
where the 1-simplices average degree ${\langle k_i \rangle}$ and 2-simplices average degree ${\langle k_i^{\Delta} \rangle}$ of individuals in the simplicial complex $\mathscr{S}_i(t)$ are used to determine the number of pairwise and higher-order interactions per day inside $i$. The term $(1-\lambda \frac{I_i^{\rm eff}(t)}{n_i^{\rm eff}})^{\langle k_i \rangle}$ indicates the probability of individuals not getting infected based on pairwise interactions in 1-simplices, while $\big(1-\lambda^{\Delta} [\frac{I_i^{\rm eff}(t)}{n_i^{\rm eff}}]^2\big)^{\langle k_i^{\Delta} \rangle}$ is interpreted as the probability that individuals not getting infected based on higher-order interactions in 2-simplices. 
The terms $n_i^{\rm eff}(t)$ and $I_i^{\rm eff}(t)$ denote the effective population and the effective number of $\mathcal{I}$-state individuals within subpopulation $i$ at time $t$ after population movements, respectively, expressed as
\begin{equation}
	\label{eq.5}
	n_i^{\rm eff}=\sum_{j=1}^N \delta_{ij}(1-p_d)n_i+p_d\frac{W_{ji}}{\sum_{l=1}^NW_{jl}}n_j,
\end{equation} 
and
\begin{equation}
	\label{eq.6}
	I_i^{\rm eff}=\sum_{j=1}^N \delta_{ij}(1-p_d)n_i\rho_i^I(t)+p_d\frac{W_{ji}}{\sum_{l=1}^NW_{jl}}n_j\rho_j^I(t),
\end{equation} 
where $\delta_{ij}=1$ if $i=j$; otherwise $\delta_{ij}=0$. The Eq.~(\ref{eq.6}) illustrates that regardless of the epidemiological conditions, $\mathcal{I}$-state individuals exhibit similar mobility patterns to susceptible $\mathcal{S}$-state populations.

Eqs.~(\ref{eq.2})-(\ref{eq.6}) constitute a closed set of equations encompassing the evolution of an s-SIS epidemic on a higher-order meta-population following a microscopic Markovian description. The final infection size across the system at time $t$ can be computed as $\rho^I(t)=1/N \sum_i \rho_i^I(t)$.

\subsection{Resilience of higher-order meta-population systems}\label{resi}
This section introduces a dimension-reduction approach \cite{Gao2016Universal}, mapping a multi-dimensional system to an effective one-dimensional equation. Such simplification enables us to apply traditional resilience tools developed for low-dimensional systems to analyze multi-dimensional systems.
\subsubsection{Dimension-reduction}
As a preparatory step, redefine the traffic network $\mathbf{R}$ with entries $R_{ij}=\frac{W_{ij}}{\sum_{l=1}^NW_{il}}$, and the population mobility network $\mathbf{H}$ with elements $H_{ij}=\delta_{ij}(1-p_d)n_i+p_dR_{ij}n_i$.

Referring to Eq.~(\ref{eq.2}), and defining $\rho_{i}^I=\rho_{i}^I(t)$, we can derive the dynamical evolution equation for $\rho_{i}^I$ as
\begin{equation}
	\label{eq.7}
	\frac{d\rho_{i}^I}{dt}=-\mu\rho_{i}^I+(1-\rho_{i}^I) \Pi_{i}.
\end{equation} 
According to the network redefinition, where
\begin{equation}
	\label{eq.8}
	\Pi_{i}=(1-p_d)Q_i+p_d\sum_{i=1}^NR_{ij}Q_j.
\end{equation} 

Next, we introduce a dimension-reduction approach to analyze the evolution process of a node in the network by quantifying the average dynamics of neighbouring nodes. Define the weighted out-degree of any subpopulation (node) in the weighted directed network $\mathbf{M} \in\{\mathbf{R},\mathbf{R}^{\mathrm{T}}, \mathbf{H}\}$ as $s_j^{\mathbf{M}}=\sum_{i=1}^NM_{ji}$, and $y_j$ is a scalar associated with node $j$. Additionally, we introduce an operator denoted as 
\begin{equation}
	\label{eq.9}
	\begin{aligned}
		\left\langle y_j\right\rangle_{n n} & =\frac{\frac{1}{N} \sum_{j=1}^N s_j^\mathbf{M} y_j}{\frac{1}{N} \sum_{j=1}^N s_j^\mathbf{M}}=\frac{\boldsymbol{I}^{T} \mathbf{M} \boldsymbol{y}}{\boldsymbol{I}^\mathrm{T} \mathbf{M} \boldsymbol{I}} \\
		& =\frac{\left\langle s_j^{\mathbf{M}} y_j\right\rangle}{\left\langle s_j^{\mathbf{M}}\right\rangle}=\mathfrak{L}(\boldsymbol{y})_\mathbf{M},
	\end{aligned}
\end{equation}
which takes a vector $\boldsymbol{y}=\{y_1,y_2,\cdots,y_N\}^\mathrm{T}$ as input to provide the $\boldsymbol{y}$'s average over all nearest neighbour nodes $\left\langle y_j\right\rangle_{nn}$ in the network $\mathbf{M} \in\{\mathbf{R},\mathbf{R}^{\mathrm{T}}, \mathbf{H}\}$. And the unit vector $\boldsymbol{I}=\{1,\cdots,1\}^\mathrm{T}$.

Therefore, assuming $y_j=Q_j$, Eq.~(\ref{eq.8}) can be rewritten as
\begin{equation}
	\label{eq.10}
	\Pi_{i}=(1-p_d)Q_i+p_d s_i^{\mathbf{R}}\mathfrak{L}(\boldsymbol{Q})_{\mathbf{R}^\mathrm{T}},
\end{equation} 
where $\boldsymbol{Q}=\{Q_1,Q_2,\cdots,Q_N\}^\mathrm{T}$ and $\mathbf{R}^\mathrm{T}$ is the transpose matrix of matrix $\mathbf{R}$. Inserting Eq.~(\ref{eq.10}) into Eq.~(\ref{eq.7}), we obtain
\begin{equation}
	\label{eq.11}
	\frac{d\rho_{i}^I}{dt}=-\mu\rho_{i}^I+(1-\rho_{i}^I)[(1-p_d)Q_i+p_d s_i^{\mathbf{R}}\mathfrak{L}(\boldsymbol{Q})_{\mathbf{R}^\mathrm{T}}].
\end{equation} 
Transforming the above equations into a unified vector representation, we have
\begin{equation}
	\label{eq.12}
	\frac{d\boldsymbol{\rho^I}}{dt}=-\mu\boldsymbol{\rho^I}+(1-\boldsymbol{\rho^I})[(1-p_d)\boldsymbol{Q}+p_d \boldsymbol{s^{\mathbf{R}}}\mathfrak{L}(\boldsymbol{Q})_{\mathbf{R}^\mathrm{T}}],
\end{equation}
where $\boldsymbol{\rho^I}=\{\rho^I_1,\rho^I_2,\cdots,\rho^I_N\}^\mathrm{T}$, and $\boldsymbol{s^{\mathbf{R}}}=\{s^{\mathbf{R}}_1,s^{\mathbf{R}}_2,\cdots,s^{\mathbf{R}}_N\}^\mathrm{T}$. 
Considering the network redefinition and applying the operator $\mathfrak{L}$ on both sides of Eq.~(\ref{eq.12}), we obtain
\begin{equation}
	\label{eq.13}
	\begin{aligned}
		\frac{d\mathfrak{L}(\boldsymbol{\rho^I})_\mathbf{H}}{dt}&\approx-\mu\mathfrak{L}(\boldsymbol{\rho^I})_\mathbf{H}+[1-\mathfrak{L}(\boldsymbol{\rho^I})_\mathbf{H}]\\&[(1-p_d)\mathfrak{L}(\boldsymbol{Q})_\mathbf{H}+p_d \mathfrak{L}(\boldsymbol{s^{\mathbf{R}}})_\mathbf{H}\mathfrak{L}(\boldsymbol{Q})_{\mathbf{R}^\mathrm{T}}].
	\end{aligned}
\end{equation}

To compute $\mathfrak{L}(\boldsymbol{Q})_\mathbf{M}$, where $\mathbf{M} \in\{\mathbf{R},\mathbf{R}^{\mathrm{T}}, \mathbf{H}\}$, recalling the expression for $Q_i$ as
\begin{equation}
	\label{eq.14}
	Q_{i}=1-(1-\lambda  \frac{I_i^{\rm eff}}{n_i^{\rm eff}})^{\langle k_i \rangle}\big(1-\lambda^{\Delta} (\frac{I_i^{\rm eff}}{n_i^{\rm eff}})^2\big)^{\langle k_i^{\Delta} \rangle},
\end{equation} 
where $I_i^{\rm eff}=\sum_{j=1}^NH_{ji}\rho_j^I$ and $n_i^{\rm eff}=\sum_{j=1}^NH_{ji}$ according to the network redefinition. Similarly, utilizing the operator defined in Eq.~(\ref{eq.9}), and assuming $y_j=\rho_j^I$, $I_i^{\rm eff}$ and $n_i^{\rm eff}$ can be rewritten as
\begin{equation}
	\label{eq.15}
	I_i^{\rm eff}=s_j^{\mathbf{H}}\mathfrak{L}(\boldsymbol{\rho^I})_\mathbf{H},
\end{equation} 
and
\begin{equation}
	\label{eq.16}
	n_i^{\rm eff}=s_j^{\mathbf{H}},
\end{equation} 
respectively. So we obtain $\frac{I_i^{\rm eff}(t)}{n_i^{\rm eff}}=\mathfrak{L}(\boldsymbol{\rho^I})_\mathbf{H}$ and substitute it into Eq.~\ref{eq.14}, yielding
\begin{equation}
	\label{eq.17}
	Q_{i}=1-[1-\lambda \mathfrak{L}(\boldsymbol{\rho^I})_\mathbf{H}]^{\langle k_i \rangle} [1-\lambda^{\Delta}\mathfrak{L}^2(\boldsymbol{\rho^I})_\mathbf{M}]^{\langle k_i^{\Delta} \rangle}.
\end{equation}

Considering the effective states of the dynamical system defined by Gao et al.\cite{Gao2016Universal}, define $x_{\rm eff}^{\mathbf{M}}=\mathfrak{L}(\boldsymbol{x})_{\mathbf{M}}$, $\mathbf{M} \in\{\mathbf{R},\mathbf{R}^{\mathrm{T}}, \mathbf{H}\}$, then
\begin{equation}
	\label{eq.18}
	\rho_{\rm eff}^{\mathbf{H}}=\mathfrak{L}(\boldsymbol{\rho^I})_{\mathbf{H}}=\frac{\boldsymbol{I}^{T} \mathbf{H} \boldsymbol{\rho^I}}{\boldsymbol{I}^\mathrm{T} \mathbf{H} \boldsymbol{I}}=\frac{\left\langle s_j^{\mathbf{H}} \rho^I_j\right\rangle}{\left\langle s_j^{\mathbf{H}}\right\rangle},
\end{equation}
and
\begin{equation}
	\label{eq.19}
	Q_{\rm eff}^{\mathbf{M}}=\mathfrak{L}(\boldsymbol{Q})_{\mathbf{M}}=\frac{\boldsymbol{I}^{T} \mathbf{M} \boldsymbol{Q}}{\boldsymbol{I}^\mathrm{T} \mathbf{M} \boldsymbol{I}}=\frac{\left\langle s_j^{\mathbf{M}} Q_j\right\rangle}{\left\langle s_j^{\mathbf{M}}\right\rangle}.
\end{equation}	
Define a control parameter as $\beta_{\rm eff}^{\mathbf{M},\mathbf{N}}=\mathfrak{L}(\boldsymbol{s^{\mathbf{M}}})_{\mathbf{N}}$, where $\mathbf{M}, \mathbf{N} \in\{\mathbf{R},\mathbf{R}^{\mathrm{T}}, \mathbf{H}\}$, then 
\begin{equation}
	\label{eq.20}
	\beta_{\rm eff}^{\mathbf{R},\mathbf{H}}=\mathfrak{L}(\boldsymbol{s^{\mathbf{R}}})_{\mathbf{H}}=\frac{\boldsymbol{I}^{T} \mathbf{H} \boldsymbol{s^{\mathbf{R}}}}{\boldsymbol{I}^\mathrm{T} \mathbf{H} \boldsymbol{I}}=\frac{\left\langle s_j^{\mathbf{H}} s_j^{\mathbf{R}}\right\rangle}{\left\langle s_j^{\mathbf{H}}\right\rangle}.
\end{equation}	
Hence Eq.~(\ref{eq.13}) can be rewritten as
\begin{equation}
	\label{eq.21}
	\frac{d\rho^{\mathbf{H}}_{\rm eff}}{dt}\approx-\mu \rho^{\mathbf{H}}_{\rm eff}+(1-\rho^{\mathbf{H}}_{\rm eff})[(1-p_d)Q_{\rm eff}^{\mathbf{H}}+p_d\beta_{\rm eff}^{\mathbf{R},\mathbf{H}}Q_{\rm eff}^{\mathbf{R}^{\mathrm{T}}}].
\end{equation}
By utilizing equation $Q_i=1-[1-\lambda \rho^{\mathbf{H}}_{\rm eff}]^{\langle k_i \rangle}[1-\lambda^{\Delta}(\rho^{\mathbf{H}}_{\rm eff})^2]^{\langle k_i^{\Delta} \rangle}$ to calculate the value of $Q_{\rm eff}^{\mathbf{M}}$, and iterating $t\rightarrow \infty$ on Eq.~(\ref{eq.21}), the s-SIS process occurring in higher-order meta-populations can be fully described.

\subsubsection{Epidemic thresholds}
Near the critical point, the epidemic becomes endemic, meaning $\rho_{i}^I=\epsilon_i^I \ll 1 \forall i$. Therefore, Eq.~(\ref{eq.14}) can be linearized as
\begin{equation}
	\label{eq.22}
	Q_{i}\simeq\lambda {\langle k_i \rangle} \frac{I_i^{\rm eff}}{n_i^{\rm eff}}+\lambda^{\Delta} {\langle k_i^{\Delta} \rangle} (\frac{I_i^{\rm eff}}{n_i^{\rm eff}})^2.
\end{equation} 
Based on the definitions in Eq.~(\ref{eq.15}) and Eq.~(\ref{eq.16}), and unifying the above equation into vector form, we have
\begin{equation}
	\label{eq.23}
	\boldsymbol{Q}\simeq\lambda \boldsymbol{\langle k \rangle} \mathfrak{L}(\boldsymbol{\rho^I})_\mathbf{H}+\lambda^{\Delta} \boldsymbol{\langle k^{\Delta} \rangle} \mathfrak{L}^2(\boldsymbol{\rho^I})_\mathbf{H}.
\end{equation} 
Applying operator $\mathfrak{L}$ on both sides, one can write Eq.~(\ref{eq.23}) as
\begin{equation}
	\label{eq.24}
	\mathfrak{L}(\boldsymbol{Q})_{\mathbf{H}}\approx\lambda \mathfrak{L}(\boldsymbol{\langle k \rangle})_\mathbf{H} \mathfrak{L}(\boldsymbol{\rho^I})_\mathbf{H}+\lambda^{\Delta} \mathfrak{L}(\boldsymbol{\langle k^{\Delta} \rangle})_\mathbf
	{H} \mathfrak{L}^2(\boldsymbol{\rho^I})_\mathbf{H},
\end{equation} 
and
\begin{equation}
	\label{eq.25}
	\mathfrak{L}(\boldsymbol{Q})_{\mathbf{R}^{\mathrm{T}}}\approx\lambda \mathfrak{L}(\boldsymbol{\langle k \rangle})_{\mathbf{R}^{\mathrm{T}}} \mathfrak{L}(\boldsymbol{\rho^I})_\mathbf{H}+\lambda^{\Delta} \mathfrak{L}(\boldsymbol{\langle k^{\Delta} \rangle})_{\mathbf{R}^{\mathrm{T}}} \mathfrak{L}^2(\boldsymbol{\rho^I})_\mathbf{H}.
\end{equation} 
Similarly, considering the effective states of the system and defining the control parameter as 
\begin{equation}
	\label{eq.26}
	\beta_{\rm eff}^{\boldsymbol{\mathscr{S}^|},\mathbf{M}}=\mathfrak{L}(\boldsymbol{\langle s \rangle^{\mathscr{S}^|}})_{\mathbf{M}}=\mathfrak{L}(\boldsymbol{\langle k \rangle})_{\mathbf{M}}=\frac{\boldsymbol{I}^{T} \mathbf{H} \boldsymbol{\langle k \rangle}}{\boldsymbol{I}^\mathrm{T} \mathbf{M} \boldsymbol{I}}=\frac{\left\langle s_j^{\mathbf{M}} \langle k_j \rangle\right\rangle}{\left\langle s_j^{\mathbf{M}}\right\rangle},
\end{equation}
and
\begin{equation}
	\label{eq.27}
	\beta_{\rm eff}^{\mathbf{\mathscr{S}^{\Delta}},\mathbf{M}}=\mathfrak{L}(\boldsymbol{\langle s \rangle^\mathbf{{\mathscr{S}^{\Delta}}}})_{\mathbf{M}}=\mathfrak{L}(\boldsymbol{\langle k^{\Delta} \rangle})_{\mathbf{M}}=\frac{\boldsymbol{I}^{T} \mathbf{H} \boldsymbol{\langle k^{\Delta} \rangle}}{\boldsymbol{I}^\mathrm{T} \mathbf{M} \boldsymbol{I}}=\frac{\left\langle s_j^{\mathbf{M}} \langle k^{\Delta}_j \rangle\right\rangle}{\left\langle s_j^{\mathbf{M}}\right\rangle},
\end{equation}		
where $\boldsymbol{\langle s \rangle}^{\mathscr{S}^{\bm{|}}}=\{\langle s \rangle^{\mathscr{S}^{\bm{|}}_1},\langle s \rangle^{\mathscr{S}^{\bm{|}}_2},\cdots,\langle s \rangle^{\mathscr{S}^{\bm{|}}_N}\}$ and $\mathbf{M} \in\{\mathbf{R}^{\mathrm{T}}, \mathbf{H}\}$. The symbols $\mathscr{S}^|_i$ and $\mathscr{S}^{\Delta}_i (i=\{1,\cdots,N\})$ denote the underlying contact networks composed of 1-simplices and 2-simplices, respectively, within the simplicial comple $\mathscr{S}_i$.  Thus, referring to Eq.~(\ref{eq.19}), Eq.~(\ref{eq.24}) and Eq.~(\ref{eq.25}) can be respectively rewritten as
\begin{equation}
	\label{eq.28}
	\boldsymbol{Q}_{\rm eff}^{\mathbf{H}}\approx\ \lambda \beta_{\rm eff}^{\boldsymbol{\mathscr{S}^|},\mathbf{H}} \rho_{\rm eff}^{\mathbf{H}}+\lambda^{\Delta} \beta_{\rm eff}^{\mathbf{\mathscr{S}^{\Delta}},\mathbf{H}}(\rho_{\rm eff}^{\mathbf{H}})^2,
\end{equation} 
and
\begin{equation}
	\label{eq.29}
	\boldsymbol{Q}_{\rm eff}^{\mathbf{R}^{\mathrm{T}}}\approx\ \lambda \beta_{\rm eff}^{\boldsymbol{\mathscr{S}^|},\mathbf{R}^{\mathrm{T}}} \rho_{\rm eff}^{\mathbf{H}}+\lambda^{\Delta} \beta_{\rm eff}^{\mathbf{\mathscr{S}^{\Delta}},\mathbf{R}^{\mathrm{T}}}(\rho_{\rm eff}^{\mathbf{H}})^2.
\end{equation}
Substituting Eq.~(\ref{eq.28}) and Eq.~(\ref{eq.29}) into Eq.~(\ref{eq.21}), we can obtain
\begin{equation}
	\label{eq.30}
	\begin{aligned}
		\frac{d\rho^{\mathbf{H}}_{\rm eff}}{dt}\approx &-\mu \rho^{\mathbf{H}}_{\rm eff}+(1-\rho^{\mathbf{H}}_{\rm eff})\bigg((1-p_d)[\lambda \beta_{\rm eff}^{\boldsymbol{\mathscr{S}^|},\mathbf{H}} \rho_{\rm eff}^{\mathbf{H}}\\&+\lambda^{\Delta} \beta_{\rm eff}^{\mathbf{\mathscr{S}^{\Delta}},\mathbf{H}}(\rho_{\rm eff}^{\mathbf{H}})^2]^{\mathbf{H}}+p_d\beta_{\rm eff}^{\mathbf{R},\mathbf{H}}\\&[\lambda \beta_{\rm eff}^{\boldsymbol{\mathscr{S}^|},\mathbf{R}^{\mathrm{T}}} \rho_{\rm eff}^{\mathbf{H}}+\lambda^{\Delta} \beta_{\rm eff}^{\mathbf{\mathscr{S}^{\Delta}},\mathbf{R}^{\mathrm{T}}}(\rho_{\rm eff}^{\mathbf{H}})^2]^{\mathbf{R}^{\mathrm{T}}}\bigg).
	\end{aligned}
\end{equation}
In Eq.~(\ref{eq.30}), the $N^2$ parameters of the microscopic description of the weighted directed network $M_{ij}$ ($M \in {R, H}$) collapse into a single macroscopic resilience parameter $\beta_{\rm eff}^{\mathbf{M},\mathbf{N}}$. In this way, the impact of any perturbation detail at the microscopic level on the state of the system can be fully represented by the corresponding changes in $\beta_{\rm eff}^{\mathbf{M},\mathbf{N}}$.

Generally, the loss of system resilience is captured by the transition from a desired to an undesired stable fixed point. Clearly, analyzing the transition of fixed points in the one-dimensional system described in Eq.(\ref{eq.30}) by characterizing the macroscopic resilience parameter $\beta_{\rm eff}^{\mathbf{M},\mathbf{N}}$, the loss of resilience described in the $N$-dimensional system by Eq.(\ref{eq.2})-Eq.~(\ref{eq.6}) can be recapitulated.

As $t\rightarrow \infty$ that is, when the system is in a steady state, it follows that $\frac{d\rho^{\mathbf{H}}_{\rm eff}}{dt}=0$. Consequently, Eq.~(\ref{eq.30}) can be rewritten as
\begin{equation}
	\label{eq.31}
	\begin{aligned}
		\mathcal{F}(\rho^{\mathbf{H}}_{\rm eff})=-\rho^{\mathbf{H}}_{\rm eff}(\rho^{\mathbf{H}}_{\rm eff}-\rho^{\mathbf{H}*}_{eff,2+})(\rho^{\mathbf{H}}_{\rm eff}-\rho^{\mathbf{H}*}_{eff,2-})
	\end{aligned}
\end{equation}
Therefore, the fixed points $\rho^{\mathbf{H}*}_{eff,1}=0$ and $\rho^{\mathbf{H}*}_{eff,2\pm}=\frac{-\mathcal{B} \pm \sqrt{\mathcal{B}^2-4\mathcal{A}\mathcal{C}}}{2 \mathcal{A}}$ can be obtained. The expressions for $\mathcal{A}$, $\mathcal{B}$ and $\mathcal{C}$ are respectively written as
\begin{equation}
	\begin{aligned}
		\mathcal{A}=(1-p_d)\lambda^{\Delta} \beta_{\rm eff}^{\mathbf{\mathscr{S}^{\Delta}},\mathbf{H}}+p_d \lambda^{\Delta} \beta_{\rm eff}^{\mathbf{R},\mathbf{H}} \beta_{\rm eff}^{\mathbf{\mathscr{S}^{\Delta}},\mathbf{R}^{\mathrm{T}}},
	\end{aligned}
\end{equation}
\begin{equation}
	\begin{aligned}
		\mathcal{B}&=(1-p_d) (\lambda \beta_{\rm eff}^{\boldsymbol{\mathscr{S}^|},\mathbf{H}}-\lambda^{\Delta} \beta_{\rm eff}^{\mathbf{\mathscr{S}^{\Delta}},\mathbf{H}})\\&+p_d \beta_{\rm eff}^{\mathbf{R},\mathbf{H}} (\lambda \beta_{\rm eff}^{\boldsymbol{\mathscr{S}^|},\mathbf{R}^{\mathrm{T}}}-\lambda^{\Delta} \beta_{\rm eff}^{\mathbf{\mathscr{S}^{\Delta}},\mathbf{R}^{\mathrm{T}}}),
	\end{aligned}
\end{equation}
\begin{equation}
	\begin{aligned}
		\mathcal{C}=(1-p_d)\lambda \beta_{\rm eff}^{\boldsymbol{\mathscr{S}^|},\mathbf{H}}+p_d \lambda \beta_{\rm eff}^{\mathbf{R},\mathbf{H}} \beta_{\rm eff}^{\boldsymbol{\mathscr{S}^|},\mathbf{R}^{\mathrm{T}}}.
	\end{aligned}
\end{equation}

The derivative of $\mathcal{F}(\rho^{\mathbf{H}}_{\rm eff})$ with respect to $\rho^{\mathbf{H}}_{\rm eff}$ at $\rho^{\mathbf{H}}_{\rm eff}=\rho^{\mathbf{H}*}_{eff,1}=0$ is calculated as
\begin{equation}
	\begin{aligned}
		\frac{d\mathcal{F}(\rho^{\mathbf{H}}_{\rm eff})}{d\rho^{\mathbf{H}}_{\rm eff}}|_{\rho^{\mathbf{H}}_{\rm eff}=0}=(1-p_d)\lambda \beta_{\rm eff}^{\boldsymbol{\mathscr{S}^|},\mathbf{H}}+p_d \lambda \beta_{\rm eff}^{\mathbf{R},\mathbf{H}} \beta_{\rm eff}^{\boldsymbol{\mathscr{S}^|},\mathbf{R}^{\mathrm{T}}}-\mu.
	\end{aligned}
\end{equation}
So the epidemic outbreak threshold $\lambda_c$ can be calculated as
\begin{equation}
	\label{eq.36}
	\begin{aligned}
		\lambda_c =\frac{\mu}{p_d \beta_{\rm eff}^{\mathbf{R},\mathbf{H}} \beta_{\rm eff}^{\boldsymbol{\mathscr{S}^|},\mathbf{R}^{\mathrm{T}}}+(1-p_d) \beta_{\rm eff}^{\boldsymbol{\mathscr{S}^|},\mathbf{H}}},
	\end{aligned}
\end{equation}
directly characterized by the mobility rate and network structure.
When $\lambda$ transitions from $\lambda<\lambda_c$ to $\lambda>\lambda_c$, the fixed point $\rho^{\mathbf{H}*}_{eff,1}$ is no longer stable, indicating a rapid shift from one fixed point to another, reflecting a process of elasticity loss in the system.

Specifically, considering the case of a uniform network, as shown in Fig.\ref{fig0}(d) for the star-shaped network, the parameters of Eq.(\ref{eq.36}) can be calculated separately as
\begin{equation}
	\begin{aligned}
		\beta_{\rm eff}^{\mathbf{R},\mathbf{H}}=1,
	\end{aligned}
\end{equation}
\begin{equation}
	\begin{aligned}
		\beta_{\rm eff}^{\boldsymbol{\mathscr{S}^|},\mathbf{R}^{\mathrm{T}}}=\frac{x\delta \langle k_1 \rangle+[\frac{1}{x}+(1-\delta)]\sum_{i=2}^N \langle k_i \rangle}{x\delta+[\frac{1}{x}+(1-\delta)]x },
	\end{aligned}
\end{equation}
and
\begin{equation}
	\begin{aligned}
		\beta_{\rm eff}^{\boldsymbol{\mathscr{S}^|},\mathbf{H}}&=\frac{n_{max} \langle k_1 \rangle + \alpha n_{max} \sum_{i=2}^N \langle k_i \rangle}{n_{max}+\alpha n_{max} x}\\&=\frac{ \langle k_1 \rangle + \alpha \sum_{i=2}^N \langle k_i \rangle}{1+\alpha  x},
	\end{aligned}
\end{equation}
where $n_{max}$ represents the population size of the central subpopulation. The subpopulation $i=1$ corresponds to the central node, while subpopulations $i=\{2,\cdots,N\}$ correspond to the remaining $x$ leaf nodes.

\section{ Numerical validation}\label{numerical}
This section aims to validate the accuracy of the dimension-reduction approach by utilizing Monte Carlo numerical simulations, and to explore the impact of higher-order interactions within subpopulations and population flow between subpopulations on the dynamics of epidemic spread.

\begin{figure*}[h]
	\centering
	\includegraphics[scale=0.85]{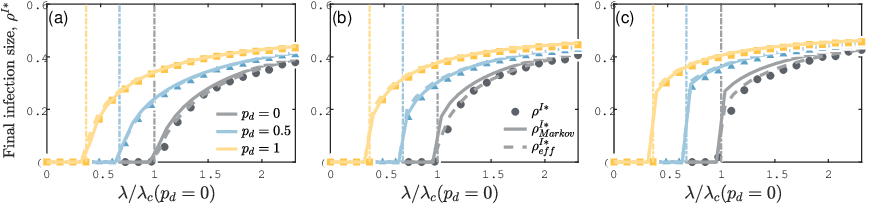}
	\caption{The final infection size $\rho^{I*}$ as a function of mobility probability $p_d$ and 1-simplices infection rate $\lambda$ with different $\lambda^{\Delta}=0.1$ (a), $\lambda^{\Delta}=0.2$ (b), and $\lambda^{\Delta}=0.3$ (c). Solid circles represent experimental results based on Monte Carlo simulations, solid lines represent theoretical results obtained using the MMCA from Eqs.(\ref{eq.2})-(\ref{eq.6}), and dashed lines represent results obtained using the dimension-reduction approach from Eq.(\ref{eq.21}). The three vertical dashed lines mark the positions of the outbreak thresholds. Different colours represent different mobility probabilities, where grey indicates no mobility $p_d=0$, blue indicates a mobility probability of $p_d=0.5$, and yellow indicates $p_d=1$. Other parameters are set as $\rho^I_i(0)=0.01$ and $\mu=1$. For clarity, $\lambda$ has been normalized to the system's epidemic threshold $\lambda_c$ at $p_d=0$.}
	\label{fig1}
\end{figure*}

\begin{figure*}[h]
	\centering
	\includegraphics[scale=0.2]{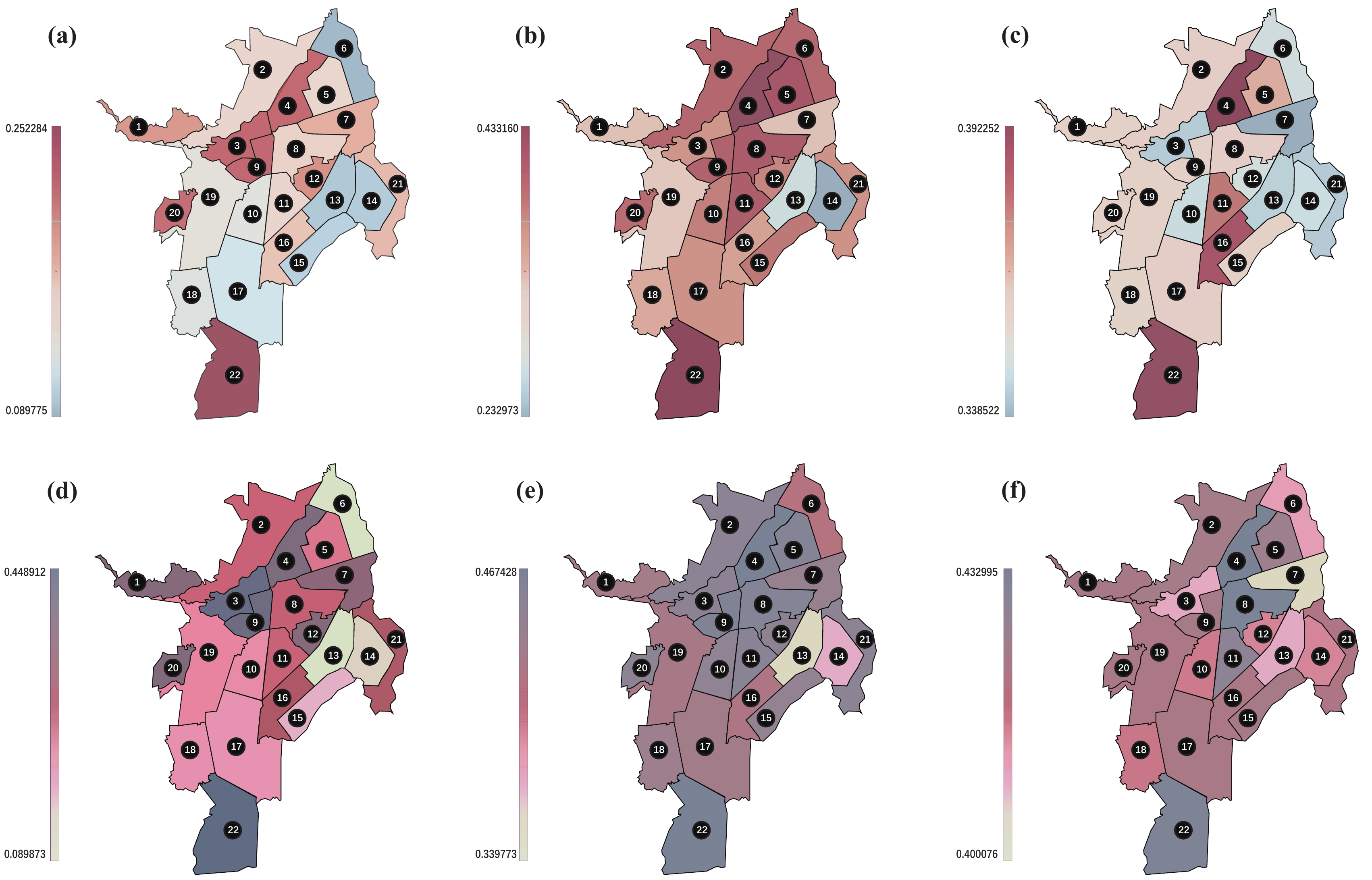}
	\caption{The spatial distribution of the epidemic in the city of Cali. Figs.~(a)-(c) depict the final infection size $\rho^{I*}_{\rm eff}$ across $22$ regions with $\lambda^{\Delta}=0.1$, calculated from Eq.~(\ref{eq.21}). Figs.~(d)-(f) illustrate the final infection size $\rho^{I*}_{\rm eff}$ across 22 regions with $\lambda^{\Delta}=0.3$. The mobility probabilities are $p_d=0$ in (a)(d), $p_d=0.5$ in (b)(e), and $p_d=1$ in (c)(f). Other parameters are $\rho^I_i(0)=0.01$, $\lambda=0.03$, and $\mu=1$.}
	\label{fig2}
\end{figure*}

\begin{figure*}[htp]
	\centering
	\includegraphics[scale=0.85]{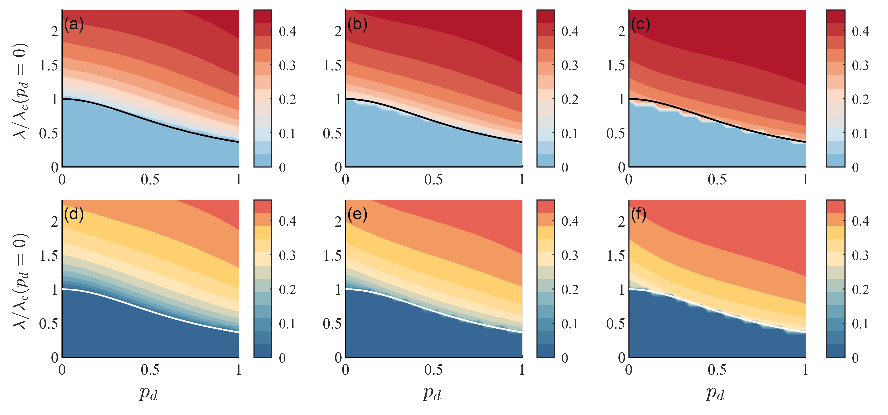}
	\caption{Phase diagram of final infection size on plane($\lambda$, $p_d$) with $\lambda^{\Delta}=0.1$ in (a)(d), $\lambda^{\Delta}=0.2$ in (b)(e), and $\lambda^{\Delta}=0.3$ in (c)(f). Figs (a)-(c) depict the variations of the solution $\rho^{I*}_{Markov}$ of equations Eqs.~(\ref{eq.2})-(\ref{eq.6}) across the parameter space. Figs (d)-(f) show the numerical experimental results for the solution $\rho^{I*}_{\rm eff}$ of Eq.(\ref{eq.21}) after dimension-reduction. The black solid line in (a-c) and the white solid line in (d-f) represent the analytical expression of $\lambda_c$ in Eq.(\ref{eq.36}). The other parameters are consistent with those in Fig.~\ref{fig1}. }
	\label{fig3}
\end{figure*}

The underlying network $\mathbf{W}$ in the first set of experiments (Fig.\ref{fig1}-Fig.\ref{fig3}) is abstracted from a real urban system, namely the city of Cali, Colombia. Referring to population distribution data obtained from the A. Arenas team, as cited in Ref.~\cite{Gomez-Gardenes2018Criticala}, Cali city comprises $2.4 \times 10^6$ permanent residents, who are formally divided into $N=22$ districts (subpopulations) based on their residential locations. The matrix $\mathbf{W}$ is constructed according to the population flow patterns between these $N=22$ subpopulations. In these experiments, the average degree of $\mathcal{S}_i$, reflecting the pairwise and  higher-order interaction patterns within subpopulation $i$, is respectively fixed as $\langle k_i \rangle=\frac{n_i^{\rm eff}}{a_i}$ and $\langle k_i^{\Delta} \rangle = \frac{\langle k_i \rangle-(n_i-1) p_1 }{2(1-p_1)}$ with $p_1 \approx \frac{\langle k_i \rangle}{n_i}$. Here, $a_i\approx \frac{n_i}{\rho_{aver}}$ is approximately calculated as the area of subpopulation $i$, where the average population density of Cali $\rho_{aver} \approx 3850/km^2$.

First, focusing on Fig.~\ref{fig1}, we conduct Monte Carlo simulations on the reaction-diffusion process with higher-order interactions based on analysing population movement in a real meta-population network. The simulation results of $\rho^{I*}$ are compared with the solutions $\rho^{I*}_{Markov}$ of Eqs.(\ref{eq.2})-(\ref{eq.6}) and the solution $\rho^{I*}_{\rm eff}$ of Eq.(\ref{eq.21}) in Fig.~\ref{fig1}. The perfect agreement between the results confirms the utility of the MMCA and the dimension-reduction approach in predicting the dynamics of epidemic spread. Furthermore, the outbreak threshold $\lambda_c$ calculated from Eq.(\ref{eq.36}) also perfectly matches the simulation results, validating the correctness of the theoretical threshold expression based on the dimension-reduction approach. Focusing on the impact of the infection rate, we observe that when the 2-simplices infection rate $\lambda^{\Delta}$ is low, as in Fig.~\ref{fig1}(a), the final infection size $\rho^{I*}$ gradually increases with the increase of the 1-simplices infection rate $\lambda$. However, as the $\lambda^{\Delta}$ increases, the rising trend of the $\rho^{I*}$ is no longer gradual but exhibits an abrupt transition, as shown in Fig. Figs.~\ref{fig1}(a) and (b). This demonstrates that higher-order interactions among individuals may lead to an explosive growth in the epidemic infection size. The experimental results also demonstrate that an increase in mobility probability $p_d$ further exacerbates the spread of the epidemic. A higher mobility probability provides individuals between different subpopulations with greater opportunities for contact, leading to an expansion of the final infection size and a decrease in the outbreak threshold.

Next, to provide a more intuitive and detailed perspective, as shown in Fig.~\ref{fig2}, we plot the infection size across the 22 regions based on specific parameter values. In Fig.~\ref{fig2}(a)-(c), we observe that as the mobility probability $p_d$ increases, besides the overall increase in the infection size across Cali, there are also certain variations in the relative infection rates among the $22$ regions. Specifically, when the population is entirely at home, i.e., $p_d=0$, regions $22$, $3$, $9$, and $4$ exhibit larger final infection sizes compared to other regions, while regions $6$, $13$, and $14$ show lower infection sizes. When the population starts to move, i.e., $p_d\neq0$, regions $16$, $8$, $11$, and $5$ experience larger final infection sizes compared to other regions, whereas regions $7$, $21$, and $3$ experience smaller final infection sizes compared to other regions. Similar phenomena are also observed in Figs.~\ref{fig2}(d)-(f). This phenomenon confirms that population mobility is a significant factor leading to changes in the spatial distribution of disease outbreaks across different regions. The influx of migrants into an area can introduce new viruses, expanding the infection size. Conversely, the mobility of susceptible individuals may result in a reduction in 
the spread of the disease locally. Besides, by increasing the 2-simplices infection rate $\lambda^{\Delta}$, we find that higher-order interactions among individuals significantly increase the final infection size of the entire city but have little effect on the spatial distribution of the epidemic outbreak. 

To further validate the findings, in Fig.~\ref{fig3}, we plot the phase diagrams of the final infection size in the parameter space ($\lambda$, $p_d$). The consistency between Fig.\ref{fig3}(a)-(c) and Fig.\ref{fig3}(d)-(f) clearly demonstrates the accuracy of our proposed evolutionary equations based on the dimension-reduction approach. Additionally, the outbreak threshold $\lambda_c$ calculated from Eq.(\ref{eq.36}) exhibits complete consistency with the outbreak threshold obtained from numerical iteration results and divides the phase diagram into two parts as epidemic outbreak (above the solid line) and infection disappearance (below the solid line). Focusing on the influence of population mobility $p_d$ and infection rates, we find that an increase in mobility significantly reduces the outbreak threshold and enlarges the final infection size under the same 1-simplices infection rate. Moreover, as the 2-simplices infection rate increases, the final infection size near the outbreak threshold in the phase diagram shows a stair-like pattern, indicating the occurrence of an abrupt transition. These experimental results are consistent with the findings in Fig.\ref{fig1} and Fig.\ref{fig2}.

\begin{figure*}[h]
	\centering
	\includegraphics[scale=0.85]{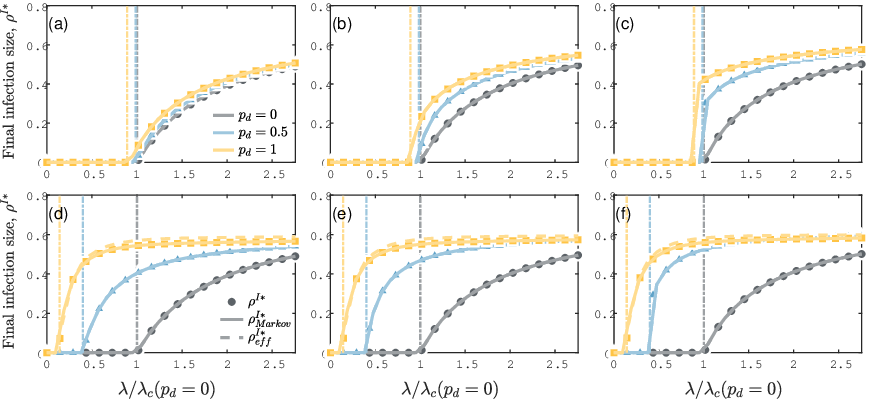}
	\caption{Reaction-diffusion processes with simplicial contagion on a synthetic star-shaped network. The meanings of different symbols and colours are consistent with those in Fig.~\ref{fig1}. The experimental results for ($\delta$, $\alpha$) = (0.2, 0.95) are shown in (a)-(c), and for ($\delta$, $\alpha$) = (0.9, 0.95) in (d)-(f). The 2-simplices infection rates are $\lambda^{\Delta} = 0.1$ in (a)(d), $\lambda^{\Delta} = 0.5$ in (b)(e), and $\lambda^{\Delta} = 0.9$ in (c)(f). Other parameters are $\rho^I_i(0) = 0.01$ and $\mu = 0.6$. The 1-simplices infection rate $\lambda$ has been rescaled by the epidemic threshold $\lambda_c$ at $p_d=0$.}
	\label{fig4}
\end{figure*}
\begin{figure*}[htp]
	\centering
	\includegraphics[scale=1.1]{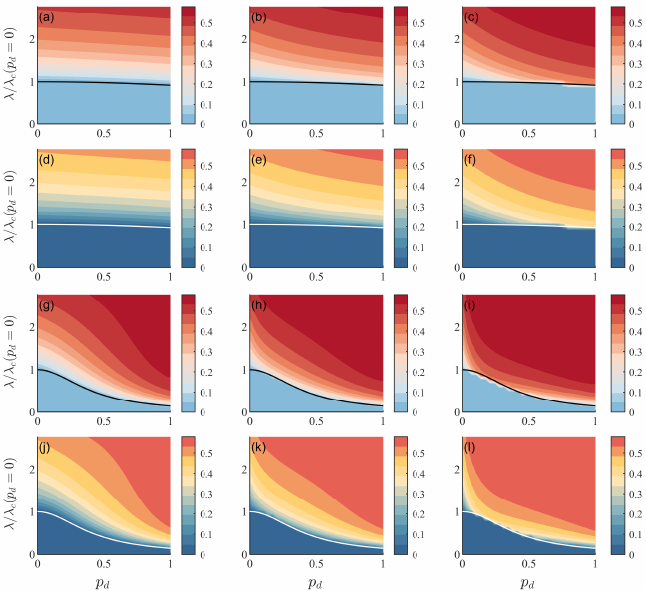}
	\caption{The phase diagrams of the final infection size in the parameter space ($\lambda$, $p_d$) based on the star network. Figs (a)-(c) and (g)-(i) show the final infection size calculated from Eqs.(\ref{eq.2})-(\ref{eq.6}), while Figs (d)-(f) and (j)-(l) display the results from Eqs.(\ref{eq.21}). In Figs (a)-(f), the parameter space ($\delta$, $\alpha$) = (0.2, 0.95), and in Figs (g)-(l), there are ($\delta$, $\alpha$) = (0.9, 0.95). The 2-simplices infection rates are $\lambda^{\Delta} = 0.1$ in (a)(d)(g)(j), $\lambda^{\Delta} = 0.5$ in (b)(e)(h)(k), and $\lambda^{\Delta} = 0.9$ in (c)(f)(i)(l). The other parameters are consistent with those in Fig.\ref{fig4}.}
	\label{fig5}
\end{figure*}

The second set of experiments (Fig.~\ref{fig4} and Fig.~\ref{fig5}) investigates a simple star-shaped network configuration consisting of a central node (subpopulation) and $k$ leaf nodes and is governed by parameters $\alpha$ and $\delta$. The parameter $\alpha$ controls the difference in population size between the central node and the leaf nodes, while $\delta$ controls the probability of population flow from the leaf nodes to the central node. In the star network, a central city acts as the central node, establishing close connections and dependencies with the surrounding regions. This central region typically possesses significant economic, cultural, or political influence, thus attracting resources, talent, and activities from neighbouring areas. In our experiment, as shown in Fig.~\ref{fig0}(d), the central node is populated by a group of size $n_{max}$, while the number of residents in each leaf node is $\alpha n_{max}$. The paths of population movement within each node follow the direction of the arrows. The mobility probability between leaf nodes is $1-\delta$, while the probability of moving from a leaf node to the central node is $\delta$. Similarly, the average degree of simplicial complexes within the star-shaped network is fixed as $\langle k_i \rangle = \frac{n_i^{\rm eff}}{a_i}$ and $\langle k_i^{\Delta} \rangle = \frac{\langle k_i \rangle - (n_i - 1) p_1}{2(1 - p_1)}$. The average population density of the star-shaped network is approximated as $\rho_{aver} = 827/km^2$, referencing the population density of Hefei, a typical representative of a star-shaped urban spatial structure.

In Fig.~\ref{fig4}, we replicate the experiments from Fig.~\ref{fig1} based on the star-shaped network. The theoretical equations based on the dimension-reduction approach in the star network can also accurately predict the epidemic spread with higher-order interactions. Figs.~\ref{fig4}(a)-(c) shows that when the probability $\delta$ of individuals moving from leaf nodes to the central node is low, and $\lambda_{\Delta}$ is small, reducing the population mobility probability does not suppress epidemic spread. In other words, if population movement is largely confined to unidirectional flow between leaf nodes, allowing mobility does not exacerbate the infection. As $\lambda_{\Delta}$ increases, the facilitation effect of population mobility on epidemic spread becomes evident, accompanied by an abrupt transition. This indicates a significant interaction between higher-order interactions among individuals and population movement, jointly influencing the dynamics of epidemic spread. Figs.~\ref{fig4}(d)-(f) shows the final infection size when individuals in all leaf nodes tend to move towards the central node. We find that an increase in population mobility probability significantly promotes the spread of the epidemic. This is because the increased mobility probability facilitates broader contact among the population in the central node, resulting in a larger final infection size.

Finally, based on the star network, we plot the phase diagrams of the final infection size in the two-parameter ($\lambda$, $p_d$) plane with different ($\delta$, $\alpha$) values in Fig.\ref{fig5}. The high consistency between Figs.(a)-(c)((g)-(i)) and Figs.~(d)-(f)((j)-(l)) reveals the effectiveness of our dimension-reduction approach in predicting epidemic spread. Figs.\ref{fig5}(a)-(f) show the experimental results when population movement is concentrated between leaf nodes, corresponding to Figs.\ref{fig4}(a)-(c). As $p_d$ changes, the outbreak threshold (black and white solid lines) remains nearly horizontal, indicating that the population mobility probability has little impact on the outbreak threshold. As $\lambda^{\Delta}$ increases, the final infection size increases with the population mobility probability, consistent with the results in Fig.~\ref{fig4}. Figs.~\ref{fig5}(g)-(l) show the experimental results when the population in the leaf nodes concentrates on moving towards the central node. The increase in $p_d$ leads to more individuals contacting the central node, significantly lowering the outbreak threshold and increasing the infection size.

\section{Conclusion} \label{conclusion}
This work constructs a higher-order meta-population model to address the limitations of traditional meta-population models, which cannot finely describe the interaction patterns within collectives with heterogeneous contact patterns. In our model, individuals within subpopulations are no longer well-mixed but are captured by a set of time-varying simplicial complexes representing pairwise and higher-order interactions. For simplified resilience analysis, we initially describe the reaction-diffusion process with simplicial contagion using an $N$-dimensional system of equations based on the MMCA. By quantifying the average dynamics of neighbouring nodes, we decompose different network behaviours into a single universal elasticity function, thereby collapsing the $N$-dimensional system into a one-dimensional equation.

Based on Monte Carlo numerical simulations, the analysis results on real networks and star networks reveal the accuracy of the proposed MMCA and the dimension-reduction framework in predicting the epidemic dynamics on higher-order meta-population networks. Additionally, we studied the impact of higher-order interactions and population mobility on the infection size and outbreak threshold. The numerical simulation results demonstrate that higher-order interactions among individuals can lead to explosive growth in the epidemic infection size. Population movement is crucial in changing the spatial distribution of infectious diseases in different regions. A significant interplay exists between higher-order interactions and population mobility, which jointly influences the epidemic spreading dynamics.

Our research significantly simplifies the resilience analysis and prediction of multidimensional higher-order meta-population network systems. The simplification not only enhances the understanding of the behaviour of higher-order complex network systems but also provides a more efficient and universal framework that lays the theoretical foundation for assessing and enhancing the stability of these systems.

\section*{Acknowledgments}
The author would like to express special gratitude to A. Arenas' team for providing real population distribution data, which has enabled us to construct realistic meta-population networks to validate our results.


\end{document}